\documentclass[pra,showpacs,floatfix,amsmath,amsfonts,twocolumn]{revtex4-1}

\usepackage{graphicx}
\usepackage{dcolumn}
\usepackage{bm}
\usepackage{color}

\usepackage{amsthm,amscd,amsfonts,amssymb,epsfig,graphicx}

\begin{document}
\date{\today}

\title{Interaction of discrete nonlinear Schr\"odinger solitons with a linear lattice impurity}

\author{Valeriy A. Brazhnyi, Chandroth P. Jisha,  A. S. Rodrigues}

\affiliation{
Centro de F\'{\i}sica do Porto, Faculdade de Ci\^encias, Universidade do Porto, R. Campo Alegre 687, Porto 4169-007, Portugal
}

\begin{abstract}
The interaction of moving discrete solitons with a linear Gaussian defect is investigated. Solitons with profiles varying from hyperbolic secant to exponentially localized are considered such that the mobility of soliton is maintained; the condition for which is obtained.
Studies on scattering of the soliton by an attractive defect potential reveal the existence of total reflection and transmission windows which become very narrow with increasing initial soliton amplitude. Transmission regions disappear beyond the small-amplitude limit. The regions of complete reflection and partial capture correspond to the windows of the existence and nonexistence of solution of the stationary problem.  Interaction of the discrete soliton with a barrier potential is also investigated. The critical amplitude of the defect at which splitting of the soliton into two parts occurs was estimated from a balance equation. The results were confirmed through direct numerical integration of the dynamical equation showing very good agreement with the analytical prediction.
\end{abstract}


 \maketitle

\section{Introduction}
Since the first theoretical prediction \cite{Cristo88} and experimental observation of discrete solitons in photonic crystals \cite{Eisenberg98} there has been a lot of activity related to the generation, stability and dynamical properties of localized modes in discrete nonlinear systems (see reviews \cite{1} and references therein). Another physical system where a discrete setting naturally appears is Bose-Einstein condensates (BECs) embedded in 1D, 2D or 3D optical lattices \cite{experiment}. For some specific parameters of the system (relatively high amplitude of the optical lattice) the Gross-Pitaevskii equation which governs the dynamics of the continuum wave function can be approximated by a discrete nonlinear Schr\"odinger (DNLS) equation \cite{2}.
The existence of discrete solitons has been demonstrated in a wide range of other physical systems such as, for example, atomic chains with on-site cubic nonlinearities \cite{takeno}, biophysical systems \cite{Davydov}, 
and in arrays of coupled nonlinear optical wave guides \cite{Eisenberg98}. Very recently discrete gap solitons were also observed experimentally in a saturable media with alternate spacing of waveguide arrays \cite{Kanshu2012}.
  
In this work, we will focus on the propagation of discrete 1D solitons through a nonlinear lattice in the presence of a localized linear defect. The classical process of scattering of an incident wave (particle) on a localized defect in the form of a barrier/well is well known and investigated. It basically consists of three different regimes; reflection, transmission and capture depending on the ratio between kinetic energy of the incoming wave/particle and the height of the defect. However treating the same process as a quantum mechanical problem results in interesting phenomena.
Interaction of a discrete soliton with a localized defect in the integrable Ablowitz-Ladik model and DNLS equation was considered theoretically in \cite{Bishop, Kro, Konotop96, Tromb, Goodman04}. The first experimental attempt to describe the discrete soliton interaction with a localized defect in arrays of coupled waveguides revealed significant differences in the linear and nonlinear regimes of the scattering process \cite{exp_def}. A great deal of attention was paid to the study of soliton interaction with a $\delta$-like linear defect in the continuum  nonlinear Schr\"odinger (NLS) equation \cite{Cao95,Frauenkron96}. Recently  this problem regained strong interest due to the possibility of using the interaction of the matter-wave solitons with a defect in BEC to investigate different quantum effects as well as to study the  interaction of BEC with surfaces \cite{Ketterle}. Matter-wave soliton interaction with extended  defect was investigated in Ref.\cite{Sakaguchi04, 
Lee06, Cornish09, 3, Helm12} where the authors discussed the phenomenon of resonant transmission and trapping of bright solitons in the presence of a quantum well. 
Interaction of  gap solitons with a localized defect in an optical lattice \cite{4, Ahufinger07} and  trapping of discrete solitons by linear and nonlinear defects  in different discrete models \cite{Forinash94, Flach03, Stoychev2004, Louis2006, Palmero08} have also been considered. 
Scattering of a single matter-wave  soliton and two-soliton molecule in a potential well was also recently considered in \cite{Baizakov}.

The article is organized as follows: the model is introduced and discussed in section (\ref{model}) where we have also obtained the frequency condition for the mobility of the soliton, allowing us to consider both hyperbolic secant and exponentially localized solitons which are not destroyed or pinned to the lattice. Section (\ref{sca_att}) discuses the scattering process for an attractive defect potential, which is followed by a defect mode analysis in section (\ref{mode_analaysis}). Finally, splitting of  discrete soliton by a repulsive defect is considered in section (\ref{rep_defect}).

\section{Model}\label{model}

One of the basic lattice models which appeared in the last decades in various contexts of physics and biology and has been intensively studied (review \cite{1}) is the DNLS equation, which has the form
\begin{equation}
i\frac{d\psi_{n}}{dt}=\Delta_{2}\psi_{n} + V_n \psi_n+2\sigma \mid \psi_{n}
\mid ^{2}\psi_{n},
\label{DNLS_t}
\end{equation}
where $\psi_{n}\equiv \psi_{n}(t)$ is the time dependent amplitude of the individual waveguide modes when an array of waveguides is considered, and is the amplitude of the individual oscillator at site $n$ when a lattice of coupled anharmonic oscillators is considered, 
$\Delta_{2}\psi_{n}\equiv\psi_{n+1}+\psi_{n-1}-2\psi_{n}$ describes the coupling between neighboring waveguides (or lattice sites),  $\sigma=\pm 1$ is the nonlinear coefficient which accounts for the optical Kerr effect (or interparticle interaction in BEC), and the linear potential $V_n$ takes into account the presence of a localized impurity in the lattice.

In the absence of the impurity, $V_n\equiv 0$, the simplest class of discrete breathers can be obtained using the steady-state ansatz $\psi _{n}(t)=e^{i\omega t}u_{n}$, where $\omega$ corresponds to the frequency of the solution. Then the stationary amplitudes $u_{n}$, $n=0,\pm 1,\pm 2...$ satisfy the lattice equation:
\begin{equation}
\Delta _{2}u_{n}+\omega u_{n}+2\sigma \mid u _{n}\mid ^{2}u _{n}=0
\label{DNLS_w}
\end{equation}
with the condition for spatial localization $u_{n}\rightarrow 0$ as $n\rightarrow \pm \infty$.

The dispersion relation, which describes the propagation of linear waves, and can be obtained considering  plane wave solution of the form $u_n(t)\propto e^{ikn}$,  is well-known:
\begin{equation}
\omega=4\sin^2(k/2),
\label{disp}
\end{equation}
and the frequency is bounded in the interval $0<\omega<4$ (linear band). Above and below  the linear band there are two semi-infinite gaps where in the presence of nonlinearity one can excite  nonlinear localized modes. 
Such modes exist due to the balance between  dispersion and nonlinear effects.
It can be shown that 
the solutions in these semi-infinite  gaps are connected through the following  relation: if the discrete function $\{u_n\}$ is the solution from the semi-infinite gap below the linear band with $\omega<0$ and $\sigma=1$ then the function $\{(-1)^n u_n\}$ is the solution of Eq.~(\ref{DNLS_w})  for $\sigma=-1$ and $\omega>4$. Therefore in the following, we will work only with localized solutions in one of the semi-infinite gap, namely, below the linear band $\omega<0$, where $\sigma=1$.

Inside the semi-infinite gap one can consider two limiting cases allowing for an analytic description of the fundamental discrete soliton. 
In the vicinity of the linear band edge the problem can be mapped to the continuum NLS equation with onsite nonlinearity or can be obtained as the tight-binding approximation of the Gross-Pitaevskii equation in the presence of deep optical lattice. 
Essentially, that leads to the sech-type approximation of the small-amplitude (envelope) discrete soliton:
\begin{equation}
u_n\approx A \ {\rm sech}(An),
\label{usech}
\end{equation}
where $A$ and $1/A$ are the amplitude and the characteristic width of the discrete soliton, respectively. For this solution, the total power $P$ and $A$ has a linear dependence:
\begin{equation}
P=\sum_n |\psi_n|^2=2A.
\label{P}
\end{equation}

In the opposite limit, when the soliton frequency lies far from the linear band with $A\gg 1$, the solution becomes highly localized (intrinsic localized mode), and an exponential localization can be used to approximate the profile of the solution:
\begin{equation}
u_n\approx Ae^{-a|n|}.
\label{uexp}
\end{equation}

\begin{figure}
\epsfig{file=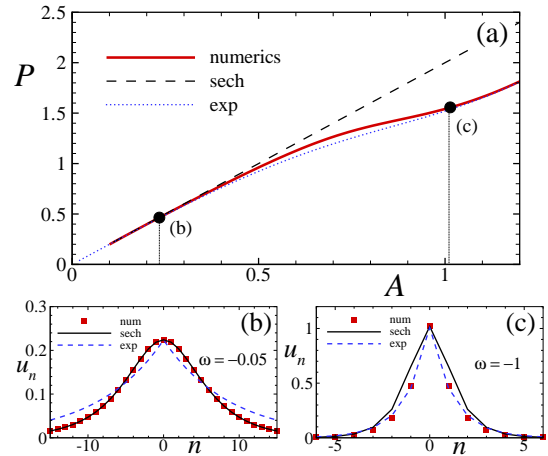,width=8cm}
\caption{(Color online) (a) Power dependence of the defect-free fundamental soliton on the central peak amplitude.  Discrete soliton profile for (b) $\omega=-0.05$ and (c) $\omega=-1$. The red boxes, black solid line and blue dashed line correspond to  numerical, sech-profile (\ref{usech}) and  exponential-profile (\ref{uexp}) solutions, respectively.} \label{fig_PA}
\end{figure}

In Fig.\ref{fig_PA}(a), numerically found  dependence of the total power $P$ of the fundamental discrete soliton on its amplitude is shown. In panels (b) and (c), the profiles of the fundamental discrete solitons calculated numerically with the two approximations (\ref{usech}) and (\ref{uexp}) are compared. As we can see, in the vicinity of the linear band the small-amplitude approximation  can be used to fit the numerical solution (see Fig.\ref{fig_PA}~(b)) while for large $|\omega|$, where $A\geq 1$, the numerical solution is well approximated by the exponential function (shown in Fig.\ref{fig_PA}~(c)).

As we are interested in the dynamical interaction of the discrete soliton  with a linear defect, it is necessary to have a stable mobile solution to observe its long-time dynamics. 
Here we discuss the limits at which the problem of interaction of discrete soliton with defect can be considered which is related to the mobility of the solitons. It is known that the mobility of  discrete solitons is restricted by the presence of the ``Peierls-Nabarro" (PN) barrier which appears due to the non-integrability of the DNLS equation \cite{PN, Peschel_2002}. 
Effectively, the PN barrier can be seen as the energy difference between on-site and inter-site fundamental modes with equal powers. Larger this difference stronger is the ``friction" between soliton and the lattice, which leads to dissipation of the soliton by radiation during propagation. At a critical amplitude of the soliton the PN barrier becomes strong enough to stop the soliton and pin it to the lattice. The mobility of discrete solitons  in terms of BEC in optical lattice under realistic experimental conditions was considered in \cite{Ahufinger04}.

\begin{figure}
\includegraphics{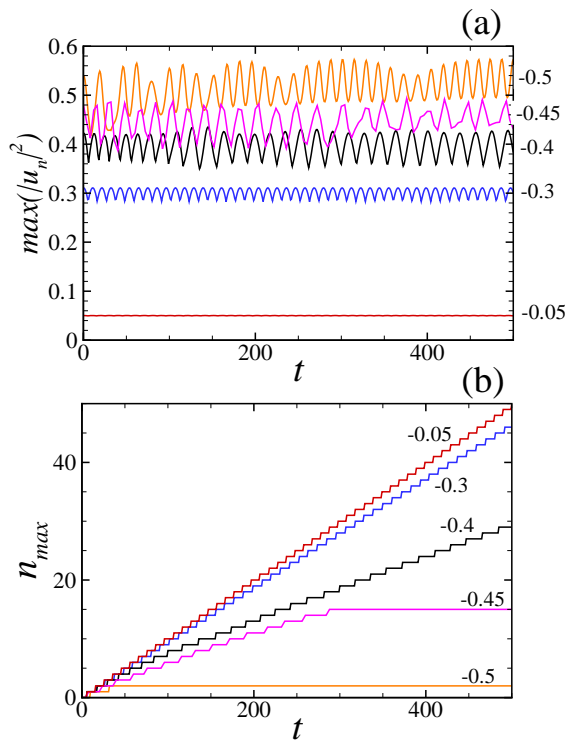}
\caption{(Color online) (a) Time dependence of the maximum amplitude $max(|u_n|^2)$ and (b) Trajectory for different frequencies $\omega=-0.05; -0.3;-0.4; -0.45; -0.5$ of the soliton with an initial velocity  $v=0.1$.  } \label{fig_PN}
\end{figure}

In order to find the frequency limit on mobility, we considered solitons with amplitude ranging from low to high, and studied their time dynamics for a fixed initial  velocity. As shown in Fig.\ref{fig_PN}, three different regimes can be observed. Small amplitude solitons propagate freely maintaining their shape and without feeling the PN barrier (for $\omega=-0.05$). Whereas, high-amplitude solitons are pinned to the lattice after some distance of propagation ($\omega=-0.45$ and $-0.5$). 
An intermediate regime is also observed, where the solitons are still able to  propagate slowing down their velocities ($\omega=-0.3$ and $-0.4$)). This implies that, considering soliton solutions in the frequency range $-0.4\leq\omega<0$, allows one to study the interaction properties of both small-amplitude and high-amplitude solitons without the soliton being dissipated or pinned.

\section{Scattering process on a potential well, $V_0>0$} \label{sca_att}

We begin by considering the interaction of the discrete soliton with an attractive defect (potential well).  As initial soliton we consider the fundamental discrete soliton of the form
\begin{equation}
\psi_n(t)=u_{n-n_s}e^{- ivn/2+i\omega_s t},
\label{phaseshift}
\end{equation}
with frequency $\omega_s$, placed far from the defect at the position $n_s$, and moving towards the impurity with a velocity $v$. In order to estimate the shift in the soliton frequency due to the applied velocity, we will use a Taylor expansion of the dispersion relation (\ref{disp}) in the vicinity of the corresponding edge of the linear band modes $k_0$
\begin{equation}
\omega(k)=\omega(k_0) +  v_B|_{k=k0}(k-k_0) + \frac{1}{2 }D|_{k=k0} (k-k_0)^2,
\label{tailor}
\end{equation}
where the Bloch velocity $v_B$ and diffraction coefficient $D$ (equivalent to the inverse effective mass in solid-state physics) are defined as follows:
\begin{equation}
v_B(k)= \frac{d\omega}{d k}= 2\sin(k), \quad D(k)= \frac{d^2\omega}{d k^2}=2\cos(k).
\label{v_BD}
\end{equation}
Notice that Eq.~(\ref{tailor}) is valid only near the stationary points $k_0$ (the bottom edge of the linear band mode corresponds to $k_0=\pm 2 \pi j$ and the top one to $k_0=\pm (2j+1) \pi $ where $j=0,1,2,...$),
where the Bloch velocity can be approximated by $v_B (k) \approx D (k-k_0)$. Therefore, by applying the initial velocity $v=v_B$ the frequency of the moving soliton $\omega_{s}(v)$ can be estimated according to the following relations:
\begin{equation}
\omega_{s}(v)=\omega_{s}+\omega_v,   \qquad \omega_v= \frac14 v^2,
\label{wsv}
\end{equation}
where we have used the fact that in the vicinity of the edge of the linear band, $k=k_0$, it follows from Eq.~(\ref{v_BD}) that $v_B(k=k_0) \equiv 0$ and $D(k=k_0) = 2$. 

The localized defect is taken in the form of a Gaussian function 
\begin{equation}
V_n=V_0 \exp\left[-(n-n_0)^2/(2d^2)\right],
\label{V}
\end{equation}
where $V_0<0$ $(V_0>0)$ and $d$ are the amplitude and the width of the barrier (well), respectively, and $n_0$ stands for the position of the impurity. In the following we will always consider the defect placed at the origin, $n_0=0$.
A Gaussian profile of the defect can be easily introduced by external beams  and can also reduce strong radiation processes \cite{4,Helm12} which appear during interaction of the soliton with boundaries of the defect with sharp profiles. 

To analyze the soliton-defect interaction, we follow the dynamics of the reflection, capture and transmission    ($RCT$) coefficients which are defined in the following way:
\begin{eqnarray}
R&=&\frac{1}{P}\sum_{n=-\infty}^{-\Delta}|\psi_{n}|^{2}\label{R}\\
C&=&\frac{1}{P}\sum_{n=-\Delta}^{\Delta}|\psi_{n}|^{2}\label{C}\\
T&=&\frac{1}{P}\sum_{n=\Delta}^{+\infty}|\psi_{n}|^{2}\label{T}
\end{eqnarray}
with the condition $R+C+T=1$. Here the normalization constant $P$ corresponds to the power of the initial stationary soliton. In order to determine the capture coefficient we take $\Delta$ equal to twice the width of the initial discrete soliton. 

\begin{figure}
\epsfig{file=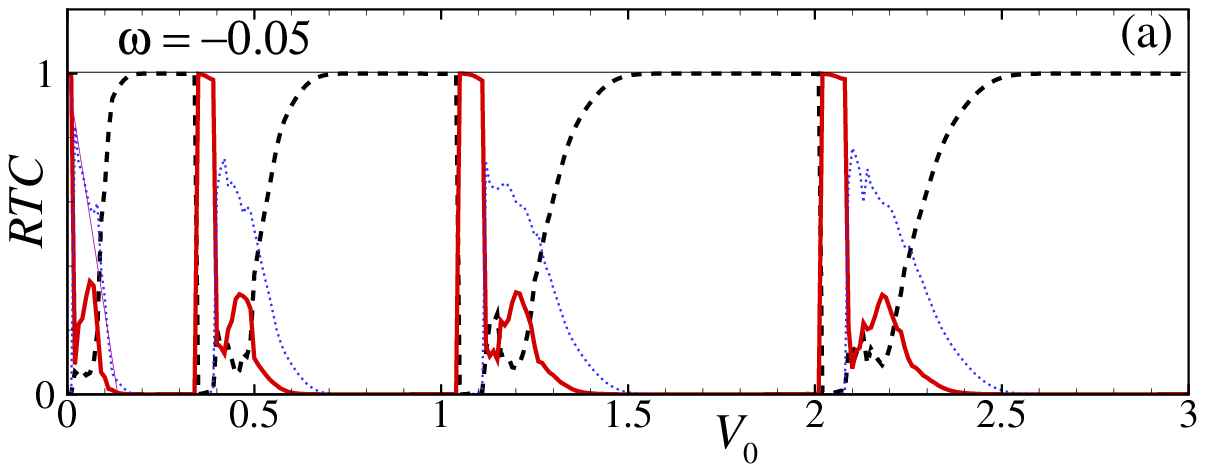,width=8cm}
\epsfig{file=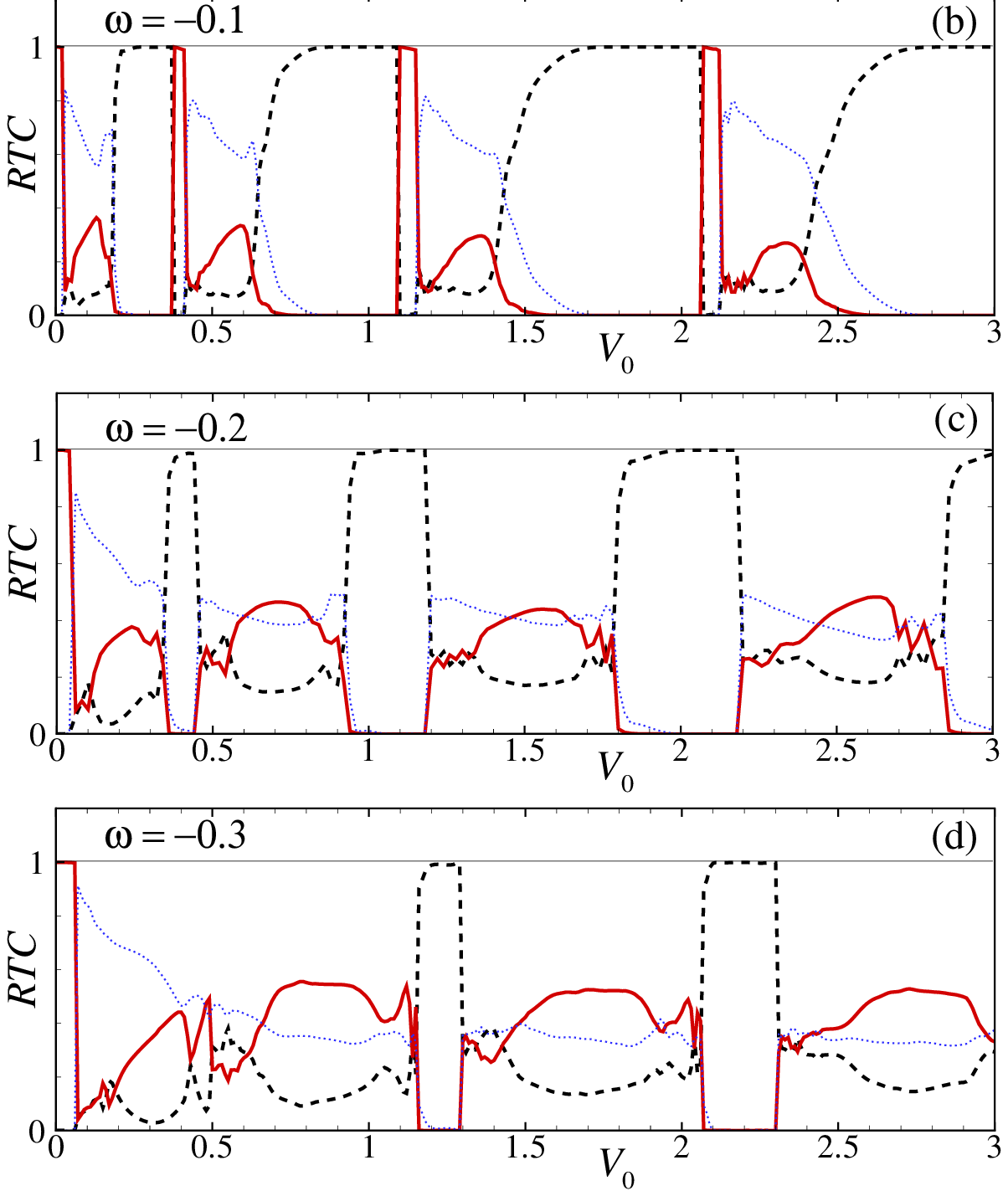,width=8cm}
 \caption{(Color online) Dependence of the reflection (dashed black line), transmission (solid red line) and  capture (blue dotted line) coefficients of the fundamental discrete soliton on the strength of the impurity, with the width $d=2$, taken at (a) $\omega_s=-0.05$ (b) $\omega_s=-0.1$ (c) $\omega_s=-0.2$ and (d) $\omega_s= -0.3$. The initial soliton velocity is $v=0.1$ and is centered at $n_s=-50$. 
} \label{TCR}
\end{figure}

\begin{figure}
\epsfig{file=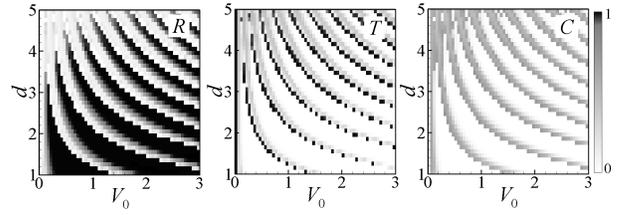,width=8cm}
\caption{Density plot for reflection (R), transmission (T) and capture (C) coefficients as a function of $(V_0, d)$ calculated as a solution of the dynamical equation with initial soliton taken at $\omega_s=-0.05$ with initial velocity $v=0.1$.  
} \label{TCR_dyn}
\end{figure}

\begin{figure}
\epsfig{file=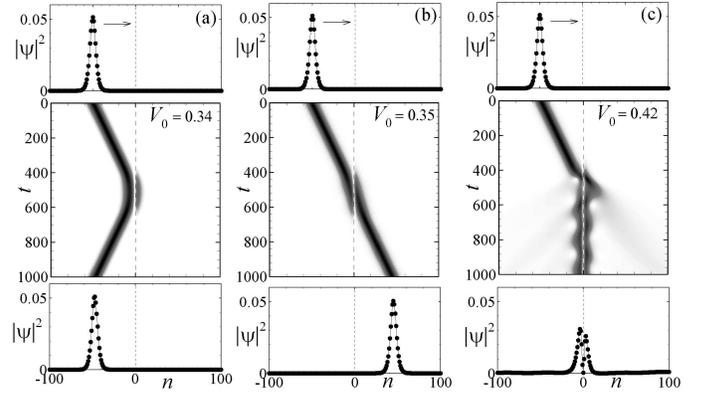,width=9cm}
\caption{Typical evolution dynamics for reflection (a), transmission (b) and capture (c). Initial and final soliton profiles are shown in top and bottom rows respectively.  
} \label{dyn}
\end{figure}

The results for the $RCT$ coefficients are obtained by the numerical integration of Eq.~(\ref{DNLS_t}). A far placed soliton with a constant velocity is impinged on to the defect and is allowed to interact with the defect for a sufficiently long time. The $RCT$ coefficients are calculated  using Eq.~(\ref{R})--(\ref{T}) and this process is repeated for different potential depths $V_0\in [0,3]$. The results are plotted in Fig.\ref{TCR}. The panels correspond to different frequencies of the initial soliton, such that the shape of the soliton changes from hyperbolic secant (low amplitude) to exponential (high amplitude) but still in the regime where the soliton has mobility and is not affected strongly by the PN barrier. The results show that for some values of the depth of the defect $V_0$ there are regions of complete reflection of the soliton by the potential well and also regions where  transmission, reflection and partial capture occur. One of the interesting features of this result is the presence of 
sharp transmission lines 
where $T\approx1$ (red line in Fig.\ref{TCR}). The dependence of $RCT$ coefficients as functions of $V_0$ and $d$ for $\omega_s=-0.05$ is plotted in Fig.\ref{TCR_dyn}.  As one can see, with increasing $d$ and $V_0$, the regions where the soliton is reflected become narrower and  windows of transmission become wider. The dynamical evolution of the small-amplitude soliton for the three different regimes; complete reflection, transmission and capture are presented in the middle row of Fig.\ref{dyn} where the top and bottom rows correspond to the initial and final profiles of the soliton respectively.

 Similar diagrams for $RCT$ coefficients on the interaction of the gap soliton with Gaussian defect in the continuous Gross-Pitaevskii equation was published recently \cite{4}. While generally the $RCT$ diagrams from Fig.\ref{TCR_dyn} shows similar structure with Fig.2 in \cite{4} the behavior of the transmission, reflection and capture curves for stronger defect are slightly different which can be explained by different definition of the defects and also presence of the optical lattice.

\begin{figure}
\epsfig{file=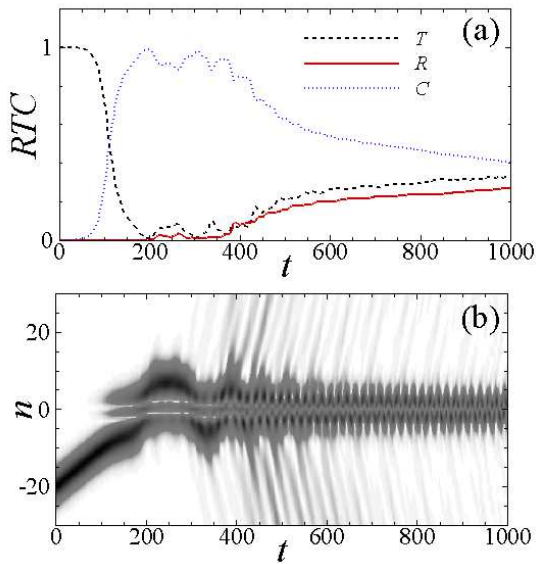,width=8cm}
\caption{(Color online) (a) Time dynamics of $RCT$ coefficients and  (b) corresponding soliton dynamics for $\omega_s=-0.2$ and $V_0=1.2$.   
} \label{TCR_dyn_w-02_120}
\end{figure}

The different frequencies in Fig.\ref{TCR} correspond to increasing amplitudes; $A=0.224$ in (a), $A=0.318$ in (b), $A=0.452$ in (c) to $A=0.558$ in (d). In all these cases the soliton is able to propagate without being destroyed or pinned to the lattice (see Fig.\ref{fig_PN}). One of the noticeable feature of the interaction of the high-amplitude soliton  with the potential well is the reduction of the width of the total reflection regions. By increasing the amplitude of the soliton one also observes that the transmission regions disappear [see Fig.\ref{TCR}(c), (d)]. This can be explained by the fact that in the high-amplitude limit the defect modes become highly unstable which prevents the discrete soliton from transforming into a stable defect mode and pass through the potential well. 
In order to show this we considered initial soliton at $\omega_s=-0.2$ and  $V_0=1.2$, where the total reflection has step-like decrease and total transmission of the discrete soliton through the defect is expected. The dynamical picture of the $RCT$ coefficients as well as evolution of the soliton profile are shown in Fig.\ref{TCR_dyn_w-02_120}. First, when the soliton enters the defect region it becomes totally captured as can be seen from the dynamics of the caption coefficient $C(t)$ at $t=200$ in Fig.\ref{TCR_dyn_w-02_120}(a). It is visible that during the passage through the defect the soliton transforms into the on-site 
defect mode with three peaks (also in agreement with the stationary defect modes analysis). However, the soliton is not able to transform into a stable mode of the defect and instead of transmission, the soliton becomes trapped by the defect, continuously radiating the extra energy and transforming itself into a one peak stable solution. This kind of unstable dynamics also explains the observed behavior of $RCT$ coefficients between reflection regions in Fig.\ref{TCR}(c), (d) where strong radiation into the transmission and reflection regions increases corresponding coefficients.

\section{Stationary defect mode analysis}\label{mode_analaysis}

\begin{figure}[h]
\epsfig{file=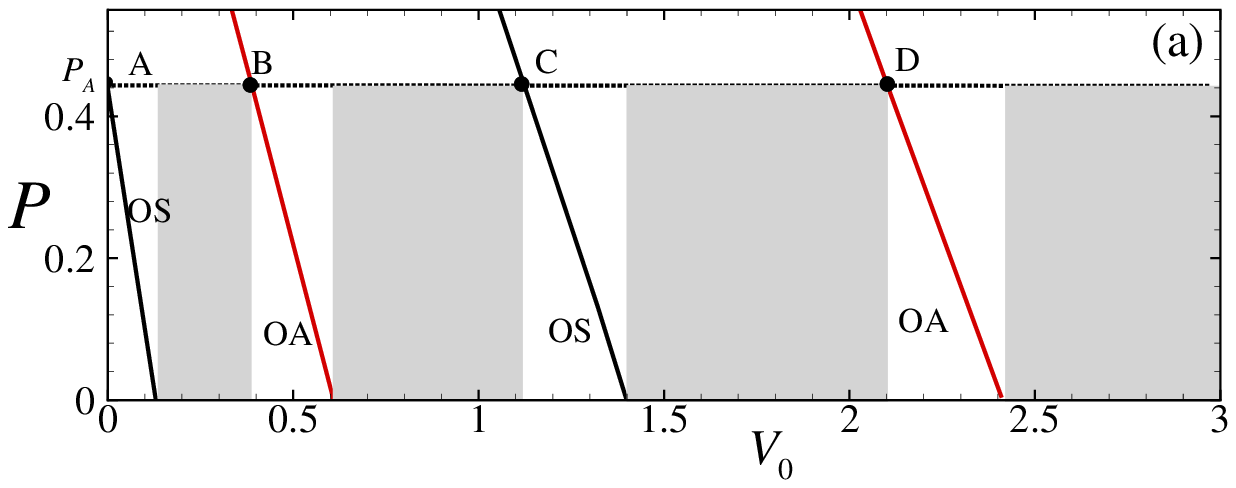,width=8cm}
\epsfig{file=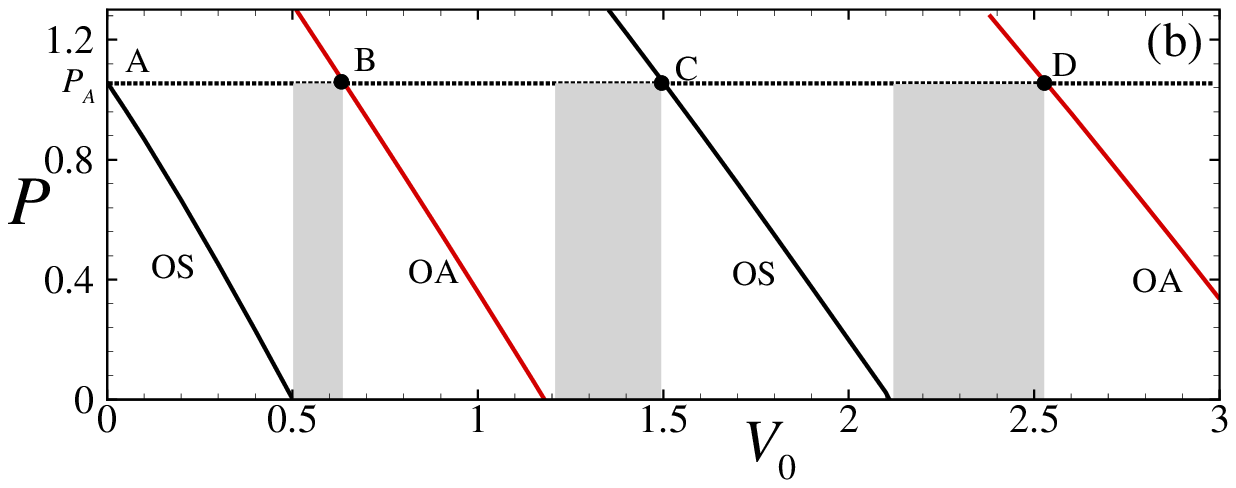,width=8cm}
 \caption{(Color online) Existence curve for the defect modes with different symmetries as a function of the strength of the potential for (a) $\omega_s=-0.05$ and (b) $\omega_s=-0.3$ for $d=2$.} \label{exist_df}
\end{figure}

\begin{figure}[h]
\epsfig{file=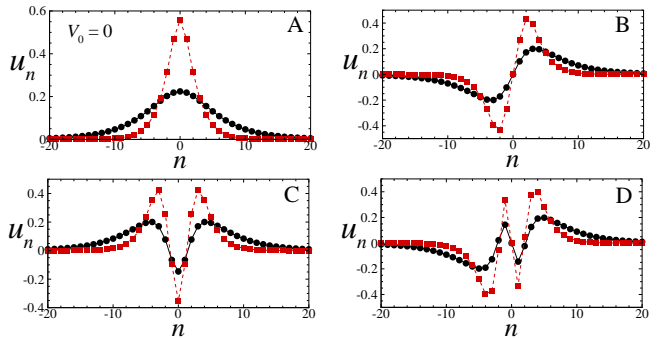,width=9cm}
 \caption{(Color online) Profiles of impurity modes at $\omega_s=-0.05$ (black solid line, circles) and $\omega_s=-0.3$ (red dashed line, squares) with $d=2$. In panels (B)-(D) the profiles of the defect modes at $V_0=0.39; 1.16; 2.1$ for $\omega=-0.05$ and $V_0=0.64; 1.5; 2.53$ for $\omega=-0.3$ with the same power as fundamental soliton $P=P_A$ presented in (A) are shown.} \label{profile_df}
\end{figure}

\begin{figure}
\epsfig{file=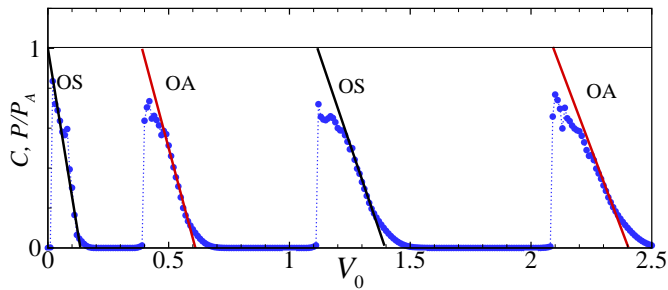,width=9cm}
 \caption{(Color online) Dependence of the capture (blue dotted line) coefficient on the amplitude of the impurity in comparison with existence curves for the SO and AO defect modes normalized to the power of the fundamental soliton ($P/P_A$). 
} \label{CC}
\end{figure}
The existence of the regions of total reflection, total transmission and partial capture of the discrete soliton by the defect can also be understood by analyzing the stationary modes of the corresponding nonlinear system \cite{4}. The stationary problem (\ref{DNLS_w}) in the presence of the linear defect now becomes
\begin{equation}
\Delta _{2}u_{n}+\omega_s u_{n} + V_n u_n + 2 \mid u _{n}\mid ^{2}u _{n}=0.
\label{DNLS_wV}
\end{equation}
Fixing the frequency of the soliton, $\omega_s$, the width of the potential well, $d$, and changing the amplitude of the potential $V_0$, we found the existence of defect modes of two different types which are dynamically stable, namely, OS (on-site symmetric) and OA (on-site antisymmetric).
The dependence of power of these defect modes with the strength of the potential with a fixed width ($d=2$) for frequencies $\omega_s=-0.05$ and $-0.3$ is presented in Fig.\ref{exist_df}.
The power of the defect-free soliton, $P_A$, is shown as the horizontal line crossing the branches of the existence of defect modes at the {\it matching points} where $P_{\rm A}=P_j$, $j=$B,C,D. In  Fig.\ref{profile_df}, the profiles of the fundamental soliton (A) and the first three defect modes at the matching points (B)-(D) are shown for both small and high-amplitude soliton. 

It is essential to mention some important characteristic features of the result shown in  Fig.\ref{exist_df}.  Considering only the defect modes whose power is less than that of the fundamental mode ($P_{\rm A}$), one can find two regions along the axis $V_0$: regions of existence of the defect modes (white boxes) and regions where there are no solutions for $P<P_{\rm A}$ (gray boxes). The width of these regions can be controlled by changing the parameters of the fundamental soliton as well as the width of the defect $d$.  As discussed earlier, interaction of the high-amplitude soliton with the defect results in the reduction in width of the region of total reflection, as can be confirmed in Fig.\ref{exist_df}(b) where increasing the soliton amplitude results in the decrease in width of the corresponding regions of non-existence of solutions. 

\begin{figure}[ht]
\epsfig{file=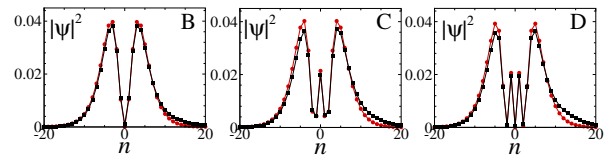,width=8cm}
 \caption{(Color online) Comparison of the profiles of the defect modes calculated from the stationary defect mode analysis (red circles) and from direct numerical simulation (black squares) at the transmission points coinciding with the points of the maximum transmission (B, C, D). 
} \label{ini_prof_num}
\end{figure}

Further analysis of the small-amplitude soliton revealed that the gray boxes exactly correspond to the intervals from the $RCT$ diagram in Fig.\ref{TCR}(a) where the discrete soliton is completely reflected and the white boxes are in excellent agreement with regions where trapping of the discrete soliton by the defect occurs. In order to confirm this, we compare in Fig.\ref{CC} the capture coefficient calculated numerically (blue line with dots)  to the branches of existence of the OS and OA defect modes (lines) calculated as steady-state solutions of Eq.~(\ref{DNLS_wV}). In order to prove that total transparency occurs at the matching points, in Fig.\ref{ini_prof_num} we compare the profiles of the defect modes, $|u_n|^2$, taken  at the points B, C and D (Fig.\ref{profile_df}) for $\omega_s=-0.05$  with snapshots of the soliton from direct numerical integration of the dynamical equation (\ref{DNLS_t}). The strength of the defect in the dynamical problem  corresponds to that of the respective matching point 
and the time of the snapshot has been chosen to show the best symmetrical situation. In all these cases complete transmission of the soliton through the defect [similar to Fig.\ref{dyn}(b)] has been observed.

\section{Splitting of the soliton by the potential barrier: $V_0<0$}\label{rep_defect}

Let us now concentrate on the interaction of the discrete soliton with a potential barrier.
Contrary to the previous case, now the soliton can only be either reflected or transmitted, depending on the relation between its initial kinetic energy and the peak energy of the barrier.
We determine the condition for the splitting of the initial soliton  into two identical parts after interaction with the impurity.
To describe the dynamics of the discrete soliton and its interaction with the linear defect including the critical conditions for the splitting, we will use two conserved quantities of the DNLS equation which are the total  power (number of particles) $P$ determined in (\ref{P}) and the Hamiltonian
\begin{eqnarray}
H&=&H_0 + H_d,\label{H}\\
H_0&=&\sum_n \left(-|\psi_n-\psi_{n-1}|^2 + \sigma|\psi_n|^4   \right), \label{H0}\\
H_d&=& \sum_n  V_n|\psi_n|^2, \label{Hd}
\end{eqnarray}
which can be rewritten in terms of the frequency in the following form:
\begin{equation}
\omega=\frac1P (H_0+H_d)=\omega_0 + \omega_d.
\label{w}
\end{equation}
Similar analysis was developed to study the scattering of bright solitons by a box-like barrier for the continuous NLS equation \cite{Sakaguchi04}.

\begin{figure}
\epsfig{file=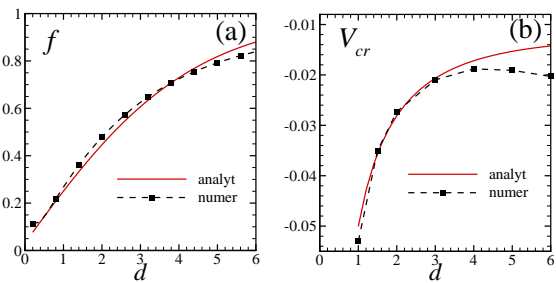,width=8cm}
 \caption{(Color online) In (a) comparison of the function $f$ calculated analytically from (\ref{f}) (red solid line) and numerically as $H_d/(2AV_0)$ (black dashed line with squares). In (b) comparison of the critical amplitude of the potential at splitting as calculated analytically from (\ref{Vcr}) (red solid line) and from direct numerical integration of the dynamical equation (black dashed line with squares). The parameters used are $A=0.224$, $\omega_s=-0.05$.
} \label{comp_f_Vcr}
\end{figure}

In order to estimate the  Hamiltonian (\ref{H}) for the single stationary soliton placed on top of the potential barrier taken in the form (\ref{V}), we used the sech-type approximation $\psi_n(t)=A\ {\rm sech}(An)e^{i\omega_st}$ for the profile of the soliton. It should be stressed here that this limit is valid only if $A\ll 1$. In this limit the total power is simply $P=2A$.
By substituting the sech-ansatz in (\ref{H}) 
the resulting Hamiltonian (frequency) becomes:
\begin{eqnarray}
H=H_0(A)+H_d(A,V_0,d)=\frac{2}{3} A^3 + 2AV_0 f(A,d),
\label{Hf}
\end{eqnarray}
where the form of the function $f(A,d)$ is determined by the width of the potential and the soliton amplitude and can be fitted with good accuracy using the following expression
\begin{equation}
f(A,d)=\tanh(Ad+A/2-0.08)
\label{f}
\end{equation}
having the two limits: as $d\to 0$ the function  $f$ tends to $A/2$ and in the limit $Ad\geq 1$ the function $f\to 1$. In Fig.\ref{comp_f_Vcr}(a) we compare the function $f$ calculated numerically using the sech-profile ansatz with the fitting function from (\ref{f}) which shows good agreement for a large range of $d$. 

By fixing the width of the potential barrier and increasing its amplitude one can find the threshold $V_{cr}$ when the discrete soliton breaks into two equal independent solutions moving from the center in opposite directions. 
The power of each part is $P'=P/2$ and the amplitude of the separated solitons is $A'=A/2$. 
Very close to the critical value the velocity of the split solitons is very small and its contribution to the total Hamiltonian is negligible. 
By neglecting this contribution we can find the critical amplitude of the potential by comparing the Hamiltonian (\ref{Hf}) with the Hamiltonian calculated for the pair of solitons far away from the defect: $H'=2H_0(A')=4/3(A')^3=A^3/6$.
Balance between these two Hamiltonians, namely, $H=H'$  gives the critical amplitude of the defect for which the splitting occurs: 
\begin{equation}
V_{cr}(A,d) = -\frac{A^2}{4f(A,d)}.
\label{Vcr}
\end{equation}
In Fig.\ref{comp_f_Vcr}(b) we compare the critical value $V_{cr}$ calculated analytically from (\ref{Vcr})  with the corresponding value obtained from the dynamical problem. 

\begin{figure}
\epsfig{file=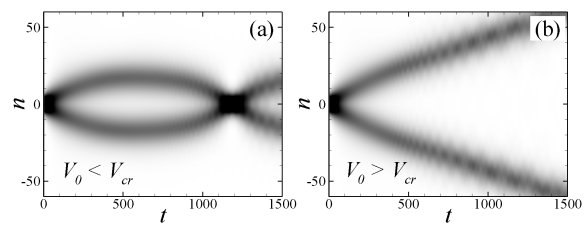,width=8cm}
 \caption{Density plot of the evolution of the soliton with $\omega_s=-0.05$ placed on top of the defect with width $d=2$ and amplitude (a) $V_0=0.027<V_{cr}$ and (b) $V_0=0.028>V_{cr}$. The critical value of the potential amplitude for these parameters is $V_{cr}=0.028$ from the analytical calculation and $V_{cr}=0.0273$ from the direct numerical calculation.
} \label{dyn_V0Vcr}
\end{figure}

Below the critical amplitude of the potential,  $V_0<V_{cr}$, one can observe breathing dynamics of the soliton placed on top of the defect. By exceeding the critical amplitude $V_0>V_{cr}$ 
the soliton splits into two solitons and extra energy coming from the difference $H-H'$ transforms into the kinetic energy (velocity) of the separated solitons. 
In Fig.\ref{dyn_V0Vcr} these two cases are presented. 

Substituting back the critical amplitude (\ref{Vcr}) into (\ref{H}) and (\ref{w}) gives us the critical energy (frequency) at which splitting occurs:
\begin{eqnarray}
H_d^{(cr)}=\frac12 A^3, \qquad \omega_d^{(cr)}= \frac14 A^2.
\end{eqnarray}

\begin{figure}
\epsfig{file=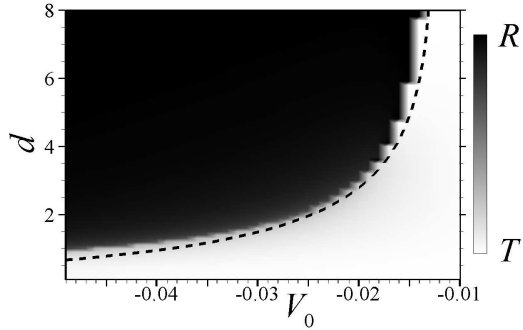,width=7cm}
 \caption{Density plot of the reflection coefficient $R$ and transmission coefficient $T$ as a function of $(V_0,d)$ with the initial soliton, taken at $\omega_s=-0.05$, placed at $n_s(t=0) = -100$ and with initial velocity $v=v_{cr}=A$. The black dashed line corresponds to the  region of splitting calculated from (\ref{Vcr}).
} \label{RT_split}
\end{figure}

Returning to the problem of the scattering of the soliton by the potential barrier, and considering now a moving soliton placed far from the defect, we can again write its Hamiltonian which now has the form
\begin{eqnarray}
H=H_0+H_v,
\label{Hv}
\end{eqnarray}
where according to (\ref{wsv}) $H_v=P \omega_v=Av^2/2$. Considering the balance equation between the Hamiltonian of the moving soliton far from the defect (\ref{Hv}) and the Hamiltonian of the soliton placed on top of the defect at critical amplitude $H_{cr}=H(V_0=V_{cr})$, from (\ref{H}) we get the critical velocity condition, $v_{cr}=A$, at which the separation between the reflection and transmission regimes occurs (due to the very large sensitivity  to the initial condition it is very difficult to find numerically the point where the soliton splits into two identical parts). 
In Fig.\ref{RT_split} the result for the reflection (transmission) coefficient as a function of $(V_0,d)$ is shown. The initial soliton velocity is taken as the critical value $v=v_{cr}$ and the separation between the reflection and transmission regions (boundary between the black and white regions) is in excellent agreement with the analytic prediction (dashed line) calculated from (\ref{Vcr}).

\section{Conclusion}
We have investigated the interaction of  discrete soliton with a linear potential taken in the form of a Gaussian function for both the small-amplitude and high-amplitude regime where the solitons are still mobile. In the case of an attractive defect, we observe the existence of  windows of complete transmission or reflection as well as regions of partial capture of the soliton by the impurity. The numerical findings are explained considering the stationary problem and the defect modes of the linear defect. Each peak of the total transmission corresponds to the matching of the power of the incoming discrete soliton with the power of the corresponding defect mode calculated from the underlying nonlinear eigenvalue problem. 
Considering the interaction of the discrete soliton with a potential barrier we found analytically the critical amplitude for the strength of the potential at which the splitting of the soliton into two identical solitons occur. The numerical calculations show excellent agreement with the analytic prediction. We expect that the obtained results can be 
interesting for the construction of  precise filters for solitons as well as for controllable  switching devices in information processing in all-optical circuits.

\section*{Acknowledgments}
JCP acknowledges support from the FCT grant SFRH/BPD/77524/2011.

\end{document}